\documentclass[12pt]{iopart}
\usepackage{iopams}
\usepackage{graphicx}
\newtheorem{lemma}{Lemma}
\newtheorem{proposition}{Proposition}
\newtheorem{theorem}{Theorem}
\newtheorem{corollary}{Corollary}
\newtheorem{definition}{Definition}
\def\tr{\mathop{\rm Tr}\nolimits}
\def\div{\mathop{\rm div}\nolimits}

\def\curl{\mathop{\rm curl}\nolimits}
\def\ci{\mathop{\textrm{i}}\nolimits}
\def\curl{\mathop{\rm curl}\nolimits}
\def\diverg{\mathop{\rm div}\nolimits}

\begin{document}

\title[Aligned Weyl fields]
{Aligned electric and magnetic Weyl fields}

\author{Joan Josep Ferrando$^1$\
and Juan Antonio S\'aez$^2$}

\address{$^1$\ Departament d'Astronomia i Astrof\'{\i}sica, Universitat
de Val\`encia, E-46100 Burjassot, Val\`encia, Spain.}

\address{$^2$\ Departament de Matem\`atica Econ\`omico-Empressarial, Universitat de
Val\`encia, E-46071 Val\`encia, Spain}

\ead{joan.ferrando@uv.es; juan.a.saez@uv.es}

\begin{abstract}
We analyze the spacetimes admitting a direction for which the
relative electric and magnetic Weyl fields are aligned. We give an
invariant characterization of these metrics and study the
properties of its Debever null vectors. The directions 'observing'
aligned electric and magnetic Weyl fields are obtained for every
Petrov type. The results on the no existence of purely magnetic
solutions are extended to the wider class having homothetic
electric and magnetic Weyl fields.

\end{abstract}




\section{Introduction}

A notable number of known solutions of Einstein equations have
been obtained by imposing restrictions on the algebraic structure
of the Weyl tensor. Thus, wide families of algebraically special
solutions have been found by considering coordinates or frames
adapted to the multiple Debever direction that these spacetimes
admit. Nevertheless, there is a lack of knowledge about
algebraically general solutions, and they have usually been
obtained by considering spacetime symmetries. One way to correct
this situation is to consider subclassifications of algebraically
general spacetimes and look for solutions in every defined class.

Some classifications of type I metrics involving first order
differential Weyl concomitants have been proposed (see
\cite{fsI1a} and references therein). Nevertheless, the more basic
restrictions that can be imposed on the Weyl tensor are the
algebraic ones. Debever \cite{deb} suggested  a classification of
type I spacetimes based on the nullity of a Weyl
invariant scalar. A similar kind of conditions are
satisfied in some classes of 'degenerate' type I spacetimes
defined by McIntosh and Arianhood \cite{mcar2}. They used the
adimensional complex scalar $M = \frac{a^3}{b^2}-6$, $a= \tr{\cal
W}^2$ and $b= \tr{\cal W}^3$ being, respectively, the quadratic
and the cubic Weyl symmetric scalar invariants. The scalar $M$ is
related to the Penrose cross-ratio invariant \cite{pen} and it
governs the geometry defined by the Debever null directions: $M=0$
in Petrov type D and, in type I, $M$ is real positive or infinite
when the four Debever directions span a 3--plane \cite{mcar2}; the
case $M$ real negative occurs when the Penrose-Rindler \cite{peri}
disphenoid associated with the Debever directions has two equal
edges \cite{armc}. Elsewhere \cite{fsI} we have presented an
alternative approach to analyzing this Debever geometry using the
complex angle between the principal bivectors and the unitary
Debever bivectors. Here we show that the case $M$ negative can be
reinterpreted in terms of permutability properties with respect to
the metric tensor of a frame \cite{como} built with the Debever
null vectors. This result was presented without proof at the
Spanish Relativity Meeting 1998 \cite{fs2}.

The electric and magnetic Weyl tensors are gravitational
quantities $E$ and $B$ attached to any observer and playing an
analogous role to the electric and magnetic fields \cite{matt}
\cite{bel}. Some classes of spacetimes can be defined by imposing
that an observer exists for which the electric and magnetic parts
of the Weyl tensor satisfy some restriction. Thus, we have the
purely electric ($B=0$) and the purely magnetic ($E=0$)
spacetimes. These conditions depend on the observer and,
consequently, they are no invariant a priory. Nevertheless,
McIntosh {\it et al.} \cite{mcar} showed that the Weyl-electric
and Weyl-magnetic spacetimes admit an intrinsic characterization
in terms of some scalar invariants: $M$ must be real positive or
infinite, and  $a$ must be real, positive in the electric case and
negative in the magnetic case. Consequently, the Debever
directions of a type I purely electric and purely magnetic Weyl
tensor span a 3--plane. Elsewhere \cite{fsEM} we have generalized
the purely electric and purely magnetic concepts by considering
electric and magnetic Weyl parts with respect to an arbitrary
direction. These generalized Weyl-electric or Weyl-magnetic
spacetimes also permit the scalar $M$ to take real negative
values, the cubic scalar $b$ being real or purely imaginary,
respectively. Thus, the new classes of gravitational fields that
we have considered in \cite{fsEM} admit a partially symmetric
frame built with Debever vectors.

All the results quoted above show that the spacetimes with the
invariant $M$ being a real function have Debever directions with
special properties, and the subfamily of them with $b^2$ a real
function can be identified in terms of the electric and magnetic
Weyl fields. Is it possible to give a characterization of the
other metrics for which $M$ is a real function by using the
relative electric and magnetic fields? In this paper we give un
affirmative answer to this question by showing that the necessary
and sufficient conditions for $M$ to be real is that the spacetime
admits a (non necessarily time-like) direction for which the
electric and magnetic parts are aligned (they are proportional
tensors). This kind of condition has been already considered for
the time-like case \cite{barnes} \cite{mcar2} and we analyze here
a generic causal character. Moreover we determine for every Petrov
type all the directions (without restriction on its causal
character) for which the relative electric and magnetic fields are
aligned.

The properties imposed on the electric and magnetic Weyl fields
imply integrability conditions which some times are very
restrictive. Thus, whereas a lot of physically interesting purely
electric solutions are known, severe restrictions appear in
dealing with purely magnetic ones (see references in \cite{fsEM}
\cite{fsM}). We want to remark here that there are no vacuum
solutions with a purely magnetic type D Weyl tensor \cite{hall},
and McIntosh {\it et al.} \cite{mcar} have conjectured that a
similar restriction could take place for a wide class of type I
spacetimes.

We can find in the literature significant steps in supporting the
McIntosh {\it et al.} conjecture. It was shown for a shear-free
observer \cite{barnes} \cite{had} and, recently, Van der Bergh has
showed the conjecture provided that the observer defines a normal
congruence \cite{van} or for a freely falling observer
\cite{van2}. Moreover, the conjecture is also true under weaker
conditions on the shear and vorticity tensors that trivially hold
when the shear or the vorticity vanish \cite{fsM}. These results
are also valid for non vacuum solutions with vanishing Cotton
tensor \cite{fsM}.

It is worth pointing out that an extension of the conjecture is
known for type D metrics. Indeed, elsewhere \cite{fsD} we have
shown that not only the purely magnetic solutions are forbidden,
but also a wider class of type D solutions. More precisely, we
have shown \cite{fsD}: {\it if a spacetime with vanishing Cotton
tensor has a type D Weyl tensor with (complex) eigenvalues of
constant argument, then it is a purely electric solution}. This
means that the constant argument takes, necessarily, the values
$0$ or $\pi$.

A question arises in a natural way: is there a similar
generalization for algebraically general spacetimes?, that is, is
it possible to extend to a wider class of type I solutions the
restrictions that one has obtained for the purely magnetic ones?
This is, precisely, the extension that we show in the present
paper: when a type I metric with vanishing Cotton tensor has
homothetic electric and magnetic Weyl fields with respect to an
observer satisfying the kinematic restrictions quoted above (those
obtained in \cite{fsM}), then the spacetime is purely electric.

The article is organized as follows. In section 2 we present the
basic formalism and we define the concepts of Weyl-aligned
spacetime and Weyl-aligned direction. In section 3 we determine,
for every Petrov type, the Weyl aligned directions. In section 4
we characterize the full class of Weyl-aligned metrics
intrinsically, as well as, some specific subclasses, and we
analyze in detail the type I Weyl-aligned spacetimes by studying
the properties of the Debever null vectors. Finally, section 5 is
devoted to extend the results on the no existence of purely
magnetic solutions to the spacetimes with homothetic electric and
magnetic fields.

\section{Weyl-aligned spacetimes}
Let $(V_4,g)$ be an oriented and time-oriented spacetime of
signature $\{ -, +,+,+ \}$, and let $W$ be its Weyl tensor. We can
associate to any unitary vector field $v$ ($v^2=\epsilon$,
$\epsilon = \pm 1$) the electric and magnetic Weyl fields:
\begin{equation}
E= E[v] \equiv W(v;v) \, , \qquad \qquad B = B[v] \equiv *W(v;v)
\label{eh}
\end{equation}
where $*$ is the Hodge dual operator and we denote $W(v;v)_{\alpha
\gamma} = W_{\alpha \beta \gamma \delta} v^{\beta} v^{\delta}$.
The electric and magnetic fields (\ref{eh}) with respect to a
spacelike or timelike congruence determine the Weyl tensor fully.
This fact was pointed out years ago for the timelike case
\cite{matt, bel}, and also holds for a spacelike congruence
\cite{fsEM}. When $v$ is a null vector we can also define the
electric and magnetic fields (\ref{eh}) but, in this case, they do
not determine the Weyl tensor \cite{fsEM}. Nevertheless, here we
also consider the electric and magnetic parts with respect to a
null direction. In this work we will use the following
definitions.

\begin{definition} \label{def1}
A metric is Weyl-aligned at a point of spacetime when there is a
vector $v$ for which the relative electric and magnetic Weyl
fields are aligned at this point. Then, the angle $\sigma \in
[0,\pi[$ such that $\cos \sigma B[v] + \sin \sigma E[v] = 0 $ is
named {\it rotation index} associated with $v$.
\end{definition}

\begin{definition}  \label{def2}
We say that $v$ is a Weyl-aligned vector if the relative electric
and magnetic Weyl fields are aligned for some rotation index
$\sigma$.
\end{definition}

These definitions extend the concepts of generalized Weyl-electric
and Weyl-magnetic spacetimes and Weyl-electric and Weyl-magnetic
directions given in \cite{fsEM}. The purely electric (resp.
magnetic) case corresponds to the rotation index to take the value
$0$ (resp. $\pi/2$). On the other hand, the rotation index has the
following interpretation: the Weyl tensor $W$ writes
\begin{equation}  \label{durot}
W = \cos \sigma W_0 + \sin \sigma *W_0
\end{equation}
$W_0$ being a Weyl-like tensor which is purely electric for the
vector $v$. That is, $\sigma$ plays the role of a duality
rotation.\\

From here we work in the self-dual complex formalism. A self--dual
2--form is a complex 2--form ${\cal F}$ such that $*{\cal F}=
\textrm{i}{\cal F}$. We can associate biunivocally to every real
2--form $F$ the self-dual 2--form ${\cal
F}=\frac{1}{\sqrt{2}}(F-\textrm{i}*F)$. We here refer to a
self--dual 2--form as a {\it bivector}. The endowed metric on the
3-dimensional complex space of the bivectors is ${\cal
G}=\frac{1}{2}(G-\textrm{i} \; \eta)$, $G$ being the usual metric
on the 2--form space, $G=\frac{1}{2}g \wedge g$, ($g\wedge
g)_{\alpha \beta \mu \nu} = 2(g_{\alpha \mu }g_{\beta
\nu}-g_{\alpha \nu}g_{\beta \mu})$, and $\eta$ being the metric
volume element. A ${\cal G}$-unitary bivector ${\cal
U}=\frac{1}{\sqrt{2}}(U-\textrm{i}*U)$ corresponds to every
timelike unitary simple 2--form $U$ ($(U,*U)=0$, $(U,U)=-1$), and
${\cal H}=\frac{1}{\sqrt{2}}(H-\textrm{i}*H)$ is a null bivector
for ${\cal G}$ when $H$ is singular ($(H,H)=(H,*H)=0$).

A unitary bivector ${\cal U}$ defines a timelike 2--plane with
volume element $U$ and its orthogonal spacelike 2--plane with
volume element $*U$. We denote these {\it principal 2--planes} as
their volume element. The null directions $l_{\pm}$ in the
2--plane $U$ are the (real) eigendirections of ${\cal U}$ and they
are called {\it principal directions}. These principal directions
may be parameterized in such a way that $U= l_- \wedge l_+$. On
the other hand a null bivector ${\cal H}$ define two null {\it
fundamental 2--planes}, with volume elements $H$ and $*H$, which
cut in the unique (real) eigendirection $l$ that ${\cal H}$
admits. Just a parametrization of the null vector $l$ exists such
that it is future-pointing and $H = l \wedge e_2$, where $e_2$ is
a spacelike unitary vector orthogonal to $l$, and fixed up to
change $e_{2}\hookrightarrow{e_{2}+\mu{l}}$. With this
parametrization we name $l$ {\it fundamental vector} of ${\cal
H}$.

The algebraic classification of the Weyl tensor $W$ can be
obtained by studying the traceless linear map defined by the
self--dual Weyl tensor ${\cal W}=\frac{1}{2} (W-\textrm{i}*W)$ on
the bivectors space. We can associate to the Weyl tensor the
complex scalar invariants
\begin{equation}
\hspace{-1.5cm}  a \equiv \tr {\cal W}^{2} = \rho_1^2 + \rho_2^2 +
\rho_3^2,  \qquad b \equiv \tr {\cal W}^{3} = \rho_1^3 + \rho_2^3
+ \rho_3^3 = 3 \rho_1  \rho_2  \rho_3   \label{ab}
\end{equation}
where $\rho_i$ are the eigenvalues. It will be also useful to
consider the adimensional scalar invariant \cite{mcar, mcar2}:
\begin{equation} \label{eme}
M \equiv \frac{a^3}{b^2}-6 = \frac{2 (\rho_1 - \rho_2)^2 (\rho_2 -
\rho_3)^2 (\rho_3 -  \rho_1)^2}{9\rho_1^2  \rho_2^2  \rho_3^2}
\end{equation}
The invariant $M$ is well defined for the Petrov types I, D or II
if we permit it to be infinite in the case of a type I metric with
$b=0$. In types D and II, $M$ is identically zero, and we extend
its validity by considering that it also takes the null value for
type N and type III metrics.

In terms of the invariants $a$ and $b$ the characteristic equation
reads $\, x^{3}-\frac{1}{2} ax -\frac{1}{3} b =0  \, $. Then,
Petrov-Bel classification follows taking into account both the
eigenvalue multiplicity and the degree of the minimal polynomial.
The algebraically regular case (type I) occurs when $6b^2 \not=
a^3$ and so the characteristic equation admits three different
roots. If $6b^2 = a^3 \not= 0$, there are a double root and a
simple one and the minimal polynomial distinguishes between types
D and II. Finally, if $a=b=0$ if all the roots are equal and so
zero, and the Weyl tensor is of type O, N or III, depending on the
degree of the minimal polynomial.

The electric and magnetic Weyl fields (\ref{eh}) associated to a
unitary vector field $v$ give, respectively, the real and
imaginary parts of the Petrov matrix ${\cal W}(v;v)$:
\begin{equation}
2{\cal W}(v;v) = W(v;v) - \ci *W(v;v) \equiv E[v] - \ci B[v]
\label{pm}
\end{equation}
On the other hand, there are four scalars built with the electric
and magnetic Weyl fields which are independent, up to sign, of the
unitary vector $v$ ($v^2=\epsilon$) \cite{matt, bel}. In fact they
are the real and imaginary parts of the complex scalar invariants
$a$ and $b$:
\begin{eqnarray}
a = (\tr E^2 - \tr B^2) - 2 \ci \tr (E \cdot B),
\label{aeh}   \\
b = - \epsilon[(\tr E^3 - 3 \tr (E \cdot B^2)) + \ci (\tr B^3 - 3
\tr (E^2 \cdot B))]         \label{beh}
\end{eqnarray}

\section{Weyl-aligned directions}

Here we determine for every Petrov type: (i) the conditions for
the spacetime to be Weyl-aligned, (ii) the rotation index $\sigma$
for which this condition holds, and (iii) the Weyl-aligned vectors
corresponding to every rotation index $\sigma$. We will express
these vectors in terms of $\sigma$ and the canonical frames or
other geometric elements associated with the Weyl tensor. As the
richness of these frames depends on the Petrov type \cite{fms}, we
will consider every algebraic class separately.

In every case, we summarize the spacetime geometry that the Weyl
{\it canonical bivectors} \cite{fms} determine, and we remark the
relationship with the canonical frames that they define. In order
to determine the Weyl-aligned directions we do not need to solve any
equation because we can use the results obtained in \cite{fsEM} on
the Weyl electric directions. Indeed, taking into account that
$\sigma$ gives the duality rotation (\ref{durot}), $W_0$ being
purely electric, we have the following:
\begin{lemma} \label{lemma1}
The necessary and sufficient condition for $v$ to be a
Weyl-aligned vector for $W$ with associated rotation index
$\sigma$ is $v$ to be a Weyl-electric vector for $\, W_0=\cos
\sigma W - \sin \sigma *W$, that is, $e^{-\ci \sigma} {\cal W}
(v;v)$ to be real.
\end{lemma}
Thus, we can use the results of \cite{fsEM} by changing ${\cal W}$
by $e^{-\ci \sigma} {\cal W}$ in every Petrov type.

\subsection{Type N}

In Petrov type N, a unique null bivector ${\cal H}$ exists such
that the self-dual Weyl tensor may be written \cite{fms}
\begin{equation}
{\cal W} = {\cal H} \otimes {\cal H}     \label{cann}
\end{equation}
The {\it canonical bivector} ${\cal H}$ determines the {\it
fundamental 2--planes} $H$ and $*H$ and the {\it fundamental
vector} $l$ (which determines the quadruple Debever direction) of
a type N Weyl tensor.

From (\ref{cann}), we have that $e^{-\ci \sigma} {\cal W}= \left(
e^{-\ci \frac{\sigma}{2}} \ {\cal H} \right) \otimes \left(
e^{-\ci \frac{\sigma}{2}} \ {\cal H} \right)$. Then, from lemma
\ref{lemma1}, the condition that a vector $v$ must satisfy to be
Weyl-aligned with associated rotation index $\sigma$ follows from
the Weyl-electric solutions in \cite{fsEM} replacing $H$ by
$\cos{\frac{\sigma}{2}}  H - \sin{\frac{\sigma}{2}}  *H$. Thus we
have:

\begin{proposition}
Every type N spacetime is Weyl-aligned, the rotation index
$\sigma$ being arbitrary. The Weyl-aligned vectors with associated
rotation index $\sigma$ are those on the planes $\
\cos{\frac{\sigma}{2}} \, H \, \pm \, \sin{\frac{\sigma}{2}} \,
*H$. These Weyl-aligned vectors are the fundamental vector $l$
(which satisfies $E[l]=B[l]=0$) and the other (spacelike) vectors
lying on the null 3--plane orthogonal to $l$.
\end{proposition}

\subsection{Type III}
In type III, a unitary bivector ${\cal U}$ and a null bivector
${\cal H}$ exist such that the self dual Weyl tensor may be written
\cite{fms}
\begin{equation} \label{can3}
{\cal W} = {\cal U} \stackrel{\sim}{\otimes} {\cal H}
\end{equation}
The {\it canonical null bivector} ${\cal H}$ determines the {\it
fundamental 2--planes} $H$ and $*H$ and the {\it fundamental
vector} $l$ (which determines the triple Debever direction). The
canonical bivectors ${\cal H}$ and ${\cal U}$ define an oriented
and ortochronous null real frame $\{ l, l^{\prime}, e_2 , e_3 \}$
such that $U= \pm \ l \wedge l^{\prime}$, $H= l \wedge e_2$. This
frame is characterized by $l$ to be the triple Debever direction,
$l^{\prime}$ the simple one, and $e_2$ (resp. $e_3$) to be the
intersection of the planes $*U$ and $H$ (resp. $*H$) \cite{fms}.

Now, $e^{-\ci \sigma} \, {\cal W} ={\cal U}
\stackrel{\sim}{\otimes} \left(e^{-\ci \sigma} \, {\cal H}
\right)$, and from lemma \ref{lemma1}, the condition that a vector
$v$ must satisfy to be
Weyl-aligned with associated rotation index $\sigma$ follows from the
Weyl-electric solutions in \cite{fsEM} replacing $H$ by
$\cos{\sigma}  H - \sin{\sigma}  *H$. Thus we have:
\begin{proposition}
Every type III spacetime is Weyl-aligned, the rotation index
$\sigma$ being arbitrary. The Weyl-aligned vectors with associated
rotation index $\sigma$ are the triple Debever direction $l$
(which satisfies $E[l]=B[l]=0$) and the spacelike direction $\,
\cos{\sigma}  e_3 - \sin{\sigma} e_2$.
\end{proposition}

\subsection{Type D}
The self-dual Weyl tensor of a Petrov type D spacetime takes the
canonical form
\begin{equation} \label{cand}
{\cal W} = 3 \rho {\cal U} \otimes {\cal U} + \rho {\cal G}
\end{equation}
where ${\cal U}$ is the {\it canonical bivector} and $\rho = -
\frac{b}{a}$ is the double eigenvalue \cite{fms}. Thus, in this
case, the {\it principal 2--planes} $U$ and $*U$ are outlined. The
principal directions $l_\pm$ of $U$ are the double Debever
directions that the type D admits.

We have $e^{-\ci \sigma} \  {\cal W} = 3 e^{-\ci \sigma}  \rho
{\cal U} \otimes {\cal U} +  e^{-\ci \sigma} \rho {\cal G}$ and,
taking into account lemma \ref{lemma1}, the conditions for ${\cal
W}$ to be Weyl-aligned follows from the results in \cite{fsEM} on
the purely electric type D metrics, just replacing $\rho$ by
$e^{-\ci \sigma} \rho$ and imposing this last expression to be
real. But it means that $\sigma$ must be either the argument
$\theta$ of the Weyl eigenvalue or $\theta - \pi $ (when $\theta$
is bigger than $\pi$). This way, we get

\begin{proposition}  \label{prode}
Every type D spacetime is Weyl-aligned and the rotation index is
$\sigma = \theta \ ({\rm mod} \ \pi)$, $\theta$ being the argument
of a Weyl eigenvalue. The Weyl-aligned vectors are the principal
ones, that is, $v\in U$ or $v \in *U$. The only null Weyl-aligned
directions are the double Debever  directions $l_{\pm}$. In the
2--plane $U$ there timelike and spacelike Weyl-aligned directions.
Every $v \in *U$ is a spacelike Weyl-aligned direction.
\end{proposition}

\subsection{Type II}
The self-dual Weyl tensor of a Petrov type II spacetime takes the
canonical form
\begin{equation} \label{can2}
{\cal W} = 3 \rho {\cal U} \otimes {\cal U} + \rho {\cal G} +
{\cal H} \otimes {\cal H}
\end{equation}
${\cal U}$ being the unitary eigenbivector associated to the
simple eigenvalue $\rho = - \frac{b}{a}$, and ${\cal H}$ being the
only eigendirection associated to the double eigenvalue. These
geometric elements define an oriented and orthochronous null real
frame $\{ l , l^{\prime} , e_2  , e_3 \}$ such that $U = \pm \ l
\wedge  l^{\prime}$ and $H = l \wedge e_2$. This frame is defined
by $l$ to be the double Debever direction, and $e_2$ (resp. $e_3$)
to be the intersection of the $2$--planes $*U$ and $H$ (resp.
$*H$) \cite{fms}.

From (\ref{can2}), we have that $e^{-\ci \sigma} \, {\cal W} = 3
e^{-\ci \sigma} \rho {\cal U} \otimes {\cal U} + e^{-\ci \sigma}
\rho {\cal G} + e^{-\ci \frac{\sigma}{2} } {\cal H} \otimes
e^{-\ci \frac{\sigma}{2}} {\cal H}$ and, taking into account lemma
\ref{lemma1}, the results in \cite{fsEM} can be applied just
changing $\rho$ by $e^{-\ci \sigma} \rho$ and ${\cal H}$ by
$e^{-\ci \frac{\sigma}{2}} {\cal H}$. Thus, we have:

\begin{proposition} \label{pro2e}
Every type II spacetime is Weyl-aligned and the rotation index is
$\sigma = \theta \ ({\rm mod} \ \pi)$, $\theta$ being the argument
of a Weyl eigenvalue. The Weyl-aligned vectors are the double
Debever direction $l$ and the spacelike directions $\cos
\frac{\theta}{2} e_2 +\sin \frac{\theta}{2} e_3$ and $- \sin
\frac{\theta}{2} e_2 + \cos \frac{\theta}{2} e_3 $.
\end{proposition}

\subsection{Type I}
The self--dual Weyl tensor of a type I spacetime takes the
canonical form:
\begin{equation} \label{can1}
{\cal W} = - \sum_{j=1}^3 \rho_j \ {\cal U}_j \otimes {\cal U}_j
\end{equation}
where $\{ {\cal U}_j \}$ are the unitary eigenbivectors associated
with the simple eigenvalues $\rho_j$ \cite{fms}. The {\it
canonical bivectors} ${\cal U}_i$ define six {\it principal
2--planes} $U_i$ and $*U_i$ which cut in the four orthogonal {\it
principal directions} that a type I metric admits. The unitary
principal vectors define the Weyl canonical frame frame
$\{e_{\alpha} \}$ that satisfies $U_i = e_0 \wedge e_i$
\cite{fsI}.

Thus, $e^{-\ci \sigma} {\cal W} = - \sum e^{-\ci \sigma}  \rho_j \
{\cal U}_j \otimes {\cal U}_j $ and, taking into account lemma 1,
the results of \cite{fsEM} can be applied changing $\rho_j$ by
$e^{-\ci \sigma} \rho_j$. These three complex numbers are real if,
and only if, the ratio between two eigenvalues is real or
infinity, the argument of every one being either $\sigma$ or $\pi
+ \sigma$. On the other hand, $e^{-\ci \sigma} \rho_j$ are complex
conjugated for two values of $j$ if, and only if, the two
eigenvalues $\rho_j$ have the same modulus, the argument $\theta$
of the third eigenvalue being either $\sigma$ or $\pi + \sigma$.
So, we have

\begin{proposition} \label{pro1e}
A type I spacetime is Weyl-aligned if, and only if, one of the two
following conditions hold:

(1) The ratio between every two eigenvalues is real (or infinity).

(2) Two of the eigenvalues have the same modulus.

If condition (1) holds, then the rotation index is $\sigma =
\theta \ ({\rm mod} \ \pi)$, $\theta$ being the argument of a Weyl
eigenvalue. Moreover, the Weyl-aligned directions are the timelike
Weyl principal direction $e_0$ and the spacelike Weyl principal
directions $e_i$.

If condition (2) holds and the third eigenvalue has different
modulus, say $|\rho_1| = |\rho_2| \not = |\rho_3|$, then the
rotation index is $\sigma = \theta_3 \ ({\rm mod} \ \pi)$,
$\theta_3$ being the argument of the eigenvalue $\rho_3$.
Moreover, the Weyl-aligned directions are the spacelike directions
$e_1 \pm e_2$.

If condition (2) holds and we have equimodular eigenvalues, then
there are three rotation index given by $\sigma_i = \theta_i \
({\rm mod} \ \pi)$, $\theta_i$ being the argument of every Weyl
eigenvalue. Moreover, the Weyl-aligned directions with associated
rotation index $\sigma_i$ are the spacelike directions $e_j \pm
e_k$, $i, j, k$ taking different values.
\end{proposition}

\section{Some classes of Weyl-aligned spacetimes}

Once the rotation index and the Weyl-aligned directions have been
found for an arbitrary Weyl tensor by considering the different
Petrov-Bel types, in this section we study and characterize same
classes of Weyl-aligned spacetimes. We begin by considering some
direct consequences of the results in previous section.
\begin{corollary}
If a metric is Weyl-aligned for a timelike or a null direction with
associated rotation index $\sigma$, then it is Weyl-aligned for a
spacelike direction with the same rotation index.\\[1mm]
A null direction is Weyl-aligned if, and only if, it is a multiple
Debever direction. Consequently, a spacetime is Weyl-aligned for a
null direction if, and only if, it is algebraically special.\\[1mm]
Every timelike Weyl-aligned direction is a Weyl principal direction.
\end{corollary}

On the other hand, we also recover the following result suggested
by Barnes \cite{barnes}:

\begin{proposition}
If a spacetime is Weyl-aligned for a timelike direction, then the
Weyl tensor is Petrov-Bel type I, D or O, and this direction is a
Weyl principal one.
\end{proposition}

We have also shown that an algebraically special spacetime is
always Weyl-aligned, but the more degenerate the Petrov type is,
the more richness in the number of rotation index exists. More
precisely, we have:

\begin{proposition}
Every algebraically special spacetime is Weyl-aligned.\\[1mm]
For Petrov types N and III ($a=b=0$) the rotation index $\sigma$
is arbitrary. For Petrov types D and II ($a^3=6b^2\not=0$) the
rotation index is $\sigma=\theta \ (mod \ \pi)$, $\theta$ being a
Weyl eigenvalue.
\end{proposition}

From here we will analyze in detail the Weyl-aligned type I
spacetimes and we will show that they can be characterized in
terms of the properties of their Debever directions. From this
study we will get the following invariant characterization of the
Weyl-aligned spacetimes:

\begin{theorem}
A spacetime is Weyl-aligned if, and only if, the Weyl invariant
scalar $M$ defined in (\ref{eme}) is real. Moreover: \\[1mm]
(i) $M=0$ if, and only if, the spacetime is
algebraically special. \\[1mm]
(ii) $M>0$ if, and only if, the spacetime is Petrov type I and it
is Weyl-aligned for a timelike direction; this direction is the
principal one and the metric is also Weyl-aligned for the three
spacelike principal directions. \\[1mm]
(iii) $M<0$ if, and only if, the spacetime is Petrov type I and it
is Weyl-aligned for the bisectors $e_i  \pm  e_j $ of a spacelike
principal 2-plane.
\end{theorem}

Let us now consider a Type I Weyl tensor. We have already shown
\cite{fsI} that, for every Weyl eigenvalue, say $\rho_3$, we can
consider the unitary bivectors ${\cal V}_{\epsilon}$, $\epsilon
=\pm 1$:
\begin{equation} \label{debeverun}
{\cal V}_{\epsilon} = \cos{\Omega} \ {\cal U}_1 + \epsilon \
\sin{\Omega} \ {\cal U}_2
\end{equation}
where the complex Weyl invariant $\Omega$ is given by
\begin{equation}\label{angulo}
\cos{2 \Omega} = \frac{3 \rho_3}{\rho_2 - \rho_1}
\end{equation}
The bivectors ${\cal V}_{\epsilon}$ are {\em unitary Debever
bivectors} \cite{fms} , that is, their principal directions are
the four simple Debever directions that a type I Weyl tensor
admits.

These expressions have been obtained privileging $\rho_3$. A
similar argument with the other two eigenvalues lead to other
pairs of Debever bivectors and gives us other angles $\Omega_1 $
and $\Omega_2$. These angles are not independent, and from
(\ref{angulo}) it is easy to show that
\begin{equation} \label{omegas}
\cos^2 {\Omega_1} = \frac{1}{\sin^2{\Omega}}, \qquad
\cos^2{\Omega_2} = - \tan^2{\Omega}
\end{equation}

Witting $\Omega = \phi - {\rm i} \psi $, we can calculate the
principal directions of the bivectors (\ref{debeverun}), and we
obtain the following expression for the Debever directions
\cite{fsI}:
\begin{equation}\label{debdir}
l_{\epsilon \pm} = \cosh \psi e_0 \pm \cos{\phi} e_1 \pm \epsilon
\sin{\phi} e_2 + \epsilon \sinh{\psi} e_3 , \qquad (\epsilon= \pm
1)
\end{equation}

On the other hand, taking into account (\ref{angulo}), the
invariant $M$ given in (\ref{eme}) can be expressed in terms of
$\Omega$. So, for every $M$, this expression poses a cubic
equation for $\cos {2 \Omega}$, being every solution associated
with one of the angles $\Omega_i$ quoted in (\ref{omegas}). More
precisely, we have for $k=0,1,2$:
\begin{equation} \label{omega-m}
\hspace{-2.4cm} \cos {2 \Omega} = \sqrt{3} \left\{N +
\frac{\beta}{2}(N-\ci) \left[ \beta \ e^{\frac{2 \pi k}{3} \ci} +
 e^{\frac{- 2 \pi k}{3}  \ci} \right]\right\}, \quad \beta \equiv
 \sqrt[3]{\frac{N+\ci}{N-\ci}},
\quad \displaystyle N \equiv \sqrt{\frac{6}{M}}
\end{equation}

Let go now on the Weyl-aligned type I metrics. We start with the
first subclass pointed out in Proposition \ref{pro1e}: the ratio
between every two eigenvalues is real (or infinity). This case
implies that the $M$ given in (\ref{eme}) is real positive (or
infinity), and this condition leads to $\cos 2\Omega$ be a real
function if we take into account (\ref{omega-m}). But if $\cos
2\Omega$ is real, the ratio between two eigenvalues is real (or
infinity) as a consequence of (\ref{angulo}). Thus, we have three
equivalent conditions.

On the other hand, if $\Omega = \phi - {\rm i} \psi $, $\cos
2\Omega$ is real when $\mbox{senh} \psi \ \mbox{cos} \phi \
\mbox{sen} \phi \ =0$. But if we take into account the expression
(\ref{debdir}), this condition states that the four Debever
directions are linearly dependent and they span the 3--plane
orthogonal to $e_j$, $\rho_j$ being the shortest eigenvalue
accordingly with (\ref{angulo}). We can summarize these results
that complete those of McIntosh et al. \cite{mcar2, mcar} (see
also \cite{fsI}) as:

\begin{theorem} \label{positivo}
In a type I spacetime the following statements are equivalent:
\begin{enumerate}
\item The metric is Weyl-aligned for a principal direction (and
then for every principal direction). \item $M$ is real positive or
infinite. \item $\cos{2 \Omega}$ is real. \item  The ratio between
every two eigenvalues is real (or infinite). \item The Debever
directions span a 3--plane.
\end{enumerate}
Moreover, if one of the above conditions hold, the 3-plane that
Debever directions span is orthogonal to $e_j$, $\rho_j$ being the
shortest eigenvalue. The case $M = \infty$ corresponds to $b
\equiv tr {\cal W}^3 =0$.
\end{theorem}

Let us consider the second subclass in Proposition \ref{pro1e}:
two of the eigenvalues have the same modulus. This means that the
ratio between these two eigenvalues lies on the unit circle,
$\displaystyle \frac{\rho_2}{\rho_1} =e^{i \theta}$, $\theta \in
(0,2\pi)$, and then the invariant $M$ given in (\ref{eme}) is real
negative or infinity. This condition implies that one of the
solutions in (\ref{omega-m}) is a purely imaginary function or
zero. But if $\cos 2\Omega_3$ is purely imaginary or zero, the
eigenvalues $\rho_2$ and $\rho_1$ have the same modulus as a
consequence of (\ref{angulo}). Thus, we have three equivalent
conditions and, taking into account proposition \ref{pro1e}, we
have established a similar result to the four first statements of
the previous theorem.

Now we look for a description of this case in terms of the Debever
directions. Elsewhere \cite{armc} the Penrose-Rindler \cite{pen}
disphenoid has been used for this purpose. Nevertheless we
interpret here this case in terms of permutability properties of a
frame  built with the Debever null vectors.

When the Debever directions $\{ l_a \}_{a=1}^4 $ are independent,
they become a null frame.  It is told that two vectors $\{ l_1 ,
l_2 \}$ of a null frame are permutable (or that the frame is
$P_2$) if $({l}_1,{l}_b) = (l_2 ,{l}_b) $  ($b=3,4$), that is, if
we can not distinguish between $l_1$ and $l_2$ making the product
with the other two vectors \cite{como}. A remarkable property is
that if a null frame is $P_2$, we can reparameterize the vectors
of the frame to make permutable the other two directions too, that
is, we can get a $P_2 \times P_2$ frame.  In the same way, it is
told that all the vectors are permutable (or that the frame is
$P_4$) if all the products $(l_a , l_b)$ ($a \neq b $) are equal
\cite{como}.

From
(\ref{debdir}) it is easy to show that if the Debever directions
are independent, they admit a reparametrization to a $P_2 \times
P_2 $ frame if, and only if, $\cos 2 \Omega$ is purely imaginary.
As we have seen, this means that there exists two of the
eigenvalues, say $\rho_1$ and $\rho_2$ ($\rho_1 \neq \pm \rho_2$)
having the same modulus. Then, the pair of permutable Debever
directions are those which are the principal directions of
$V_{\epsilon}$ constructed privileging $\rho_3$, that is the
principal directions of the Debever bivectors such that their
bisectors are ${\cal U}_1$ and ${\cal U}_2$.

The particular case of the three eigenvalues having the same
modulus leads to $M=-6$ ($a=0$) and the three solutions of
(\ref{omega-m}) are $\cos{\Omega_k} =\sqrt{3} \mbox{\rm i}$.
Moreover then, and only then, a reparametrization of the Debever
vectors exists such that we can built a $P_4$-frame. All these
results can be summarized as:

\begin{theorem} \label{negativo}

In a type I spacetime the following statements are equivalent:
\begin{enumerate}
\item The metric is Weyl-aligned for a direction which is not a
principal one. \item $M$ is real negative or infinity. \item
$\cos{2 \Omega}$ is purely imaginary or zero. \item There exist
two eigenvalues such that its ratio lies in the unit circle (have
the same modulus). \item The frame of Debever vectors can be
reparameterized to be $P_2 \times P_2$.
\end{enumerate}

Moreover, if $\rho_1$ and $\rho_2$ have the same modulus, the
pairs of permutable vectors are the principal directions of the
Debever bivectors given in (\ref{debeverun}).  In this case
$V_{\epsilon}$ and $*V_{-\epsilon}$ cut each other in the
bisectors $e_1 \pm e_2$ which are the Weyl-aligned directions of
point (i).

The frame of Debever vectors can be reparameterized to be $P_4$
if, and only if, all the eigenvalues have the same modulus, that
is when $M=-6$ ($a=0$). In this case the metric is Weyl-aligned
for the bisectors $e_i \pm e_j$ of every spacelike principal
plane.
\end{theorem}

\section{Homothetic electric and magnetic Weyl fields in vacuum:
kinematic restrictions}

Several results are known that restrict the existence of purely
magnetic spacetimes. From the initial one by Hall \cite{hall}
which showed that there no purely magnetic type D vacuum
solutions, some works are known that extend this result in
different ways. In one hand, the extension for Type I metrics
conjectured by McIntosh {\it et al.} \cite{mcar} has been shown
when the observer is: (i) shear-free \cite{barnes} \cite{had},
(ii) vorticity-free \cite{van}, (iii) geodesic \cite{van2}. The
vorticity-free and shear-free conditions have been weakened
recently \cite{fsM} by means of first-order differential
conditions which hold trivially when $\sigma=0$ or $\omega=0$. In
this last work another kind of progression is acquired: the
restriction is also valid for non vacuum solutions with vanishing
Cotton tensor. This extension has been also shown for type D
spacetimes in a paper \cite{fsD} where a third kind of
generalization is obtained: not only the purely magnetic solutions
are forbidden, but also those whose Weyl eigenvalue has a constant
argument different of $0$ or $\pi$.

In this section we will give a similar extension for type I
spacetimes. Indeed, as we have shown in section 3, every type D
metric is Weyl-aligned and the rotation index is given by the
argument of the Weyl eigenvalue. Thus, the extension for type D
spacetimes quoted above applies when, for an observer, the
electric and magnetic Weyl fields satisfy $E = k B$, $k$ being a
constant factor. Now we generalize the kinematic restrictions
obtained in \cite{fsM} to the type I spacetimes with this {\it
homothetic} property. We start by giving the following

\begin{definition}
We will say that the electric and magnetic Weyl fields $E$ and $B$
with respect to an observer $u$ are homothetic if they are aligned
with a constant rotation index $\sigma$, that is, $\cos \sigma B +
\sin \sigma E = 0$, ${\rm d} \sigma =0$.
\end{definition}
The case $\sigma = \pi/2$ corresponds to the purely magnetic case
which has been analyzed in \cite{fsM}. We will show now that
homothetic spacetimes are subjected to similar restrictions on the
kinematic coefficients of the observer that the purely magnetic
ones.

Under the hypothesi of a vanishing Cotton tensor, the Bianchi
identities take the same expression than in the vacuum case
\cite{fsM}. Thus they may be written in the 1+3 formalism
\cite{maar}:

\begin{enumerate}
\item $\diverg E = -3 \  B(\omega) +[  \sigma , B ] $
\item $ \diverg B = - \ [ \sigma , E ]  + 3 \  E(\omega)$
\item $\hat{\dot{E}} - \curl B = - \theta \ E + 3 \ E \hat{\times} \sigma  -
 \ \omega \wedge E + 2 \  \ a \wedge B $
\item $\hat{\dot{B}} + \curl E = - \theta B + 3 \
\sigma \hat{\times} B - \omega \wedge B - 2  \ a \wedge E $

\end{enumerate}
where $D$ is the covariant spatial derivative, $\div$ and $\curl$
are, respectively, the covariant spatial divergence and curl
operators, $\wedge$ and $[\ , \ ]$ are the generalized covariant
vector products and
 $\hat{}$ means the projected trace-free symmetric part
(see for example \cite{maar} for more details). Now, if $\sigma$
is constant and $\sigma \not=0$, then $E = -\cot \sigma B$, and
removing $E$ from the equations above, a straightforward
calculation leads to:
\begin{eqnarray}
[\sigma, B] = 3B(\omega)  \label{bi1} \\
\div B = 0   \label{bi2} \\
\curl B = - 2 a  \wedge B  \label{bi3}
\end{eqnarray}
But these are the same restrictions that we have used in
\cite{fsM} for the $E=0$ case. Then, taking into account the
results in \cite{fsM}, we can state:

\begin{theorem} \label{theo2}
In a spacetime with vanishing Cotton tensor if the electric and
magnetic fields are homothetic with respect to an observer $u$
satisfying one of the following conditions:
\begin{description}
    \item (i) $\tr(\curl \sigma)^2 - 3 \tr (\hat{D \omega} + 2 a
\widehat{ \otimes} \omega)^2 \not= 2(\curl \sigma, \hat{D \omega}
+ 2 a \widehat{ \otimes} \omega)$
    \item (ii) $\tr (\hat{D
\omega} + 2 a \widehat{ \otimes} \omega)^2=0$
    \item (iii) $\tr (\hat{D \omega} + 2 a
\widehat{ \otimes} \omega)^2 \geq \tr (\curl \sigma)^2$
\end{description}
Then, the spacetime is purely electric and $u$ is a Weyl principal
direction.
\end{theorem}
From here, a corollary follows.

\begin{corollary}
In a spacetime with vanishing Cotton tensor if the electric and
magnetic fields are homothetic with respect to a shear-free or a
vorticity-free observer $u$, then the spacetime is purely electric
and $u$ is a Weyl principal direction.
\end{corollary}

The result which states that the vacuum solutions with electric
and magnetic Weyl fields proportional for a shear-free observer
are, necessarily, purely electric has been also presented recently
by Barnes \cite{barere}.

\ack This work has been supported by the Spanish Ministerio de
Ciencia y Tecnolog\'{\i}a, project AYA2003-08739-C02-02.

\end{document}